\documentclass{iau307}
\usepackage{graphicx}
\usepackage{natbib}
\usepackage{url}
\usepackage{dtklogos}
\bibpunct{(}{)}{;}{a}{}{,}

\title[Observational techniques to constrain convection] 
{Combining observational techniques to constrain convection in evolved massive star models}

\author[C. Georgy, H. Saio, G. Meynet]   
{C. Georgy$^1$, H. Saio$^2$, \and G. Meynet$^3$
 }

\affiliation{$^1$Astrophysics group, Lennard-Jones Laboratories, EPSAM, Keele University, Staffordshire ST5 5BG, UK
  \\ email: {\tt c.georgy@keele.ac.uk}\\[\affilskip]
  $^2$Astronomical Institute, Graduate School of Science, Tohoku University, Sendai 980-8578, Japan\\[\affilskip]
  $^3$Geneva Observatory, University of Geneva, Maillettes 51, CH-1290 Versoix, Switzerland}

\pubyear{2014}
\volume{307} 
\pagerange{}
\jname{New windows on massive stars: asteroseismology, interferometry, and spectropolarimetry}
\editors{C. Georgy, G. Meynet, J.H. Groh \& Ph. Stee, eds.}

\begin{document}

\maketitle

\begin{abstract}

Recent stellar evolution computations indicate that massive stars in the range $\sim 20-30\,M_\odot$ are located in the blue supergiant (BSG) region of the Hertzsprung-Russell diagram at two different stages of their life: immediately after the main sequence (MS, group 1) and during a blueward evolution after the red supergiant phase (group 2). From the observation of the pulsationnal properties of a subgroup of variable BSGs ($\alpha$ Cyg variables), one can deduce that these stars belongs to group 2. It is however difficult to simultaneously fit the observed surface abundances and gravity for these stars, and this allows to constrain the physical processes of chemical species transport in massive stars. We will show here that the surface abundances are extremely sensitive to the physics of convection, particularly the location of the intermediate convective shell that appears at the ignition of the hydrogen shell burning after the MS. Our results show that the use of the Ledoux criterion to determine the convective regions in the stellar models leads to a better fit of the surface abundances for $\alpha$ Cyg variables than the Schwarzschild one.

\keywords{stars: abundances--stars: early-type--stars: evolution--stars: mass loss--stars: oscillations.}
\end{abstract}

\firstsection 

\section{Introduction}

The post main-sequence (MS) evolution of massive stars is still poorly understood, and the prediction of the simulations with different stellar evolution codes lead to very different results \citep{Martins2013a}, particularly due to the different way the transport mechanisms (convection, rotational mixing) are implemented. To improve our knowledge of massive stars evolution, it is thus of prime importance to find some observational tests that allow to discriminate between the various existing prescriptions for the internal transport mechanisms.

Some arguments seem to indicate that the mass-loss rates used in the stellar evolution codes during the red supergiant (RSG) phase \citep[often, the rates by][]{deJager1988a} could be underestimated \citep[see the discussions in][as well as \citealt{Vanbeveren1998a}]{Georgy2012a,Georgy2012b}. Recently, the Geneva group has released a new set of stellar models, including such an increased mass-loss rates during the RSG phase \citep{Ekstrom2012a,Georgy2013b}. These models show that for stars in the mass range $\sim 20-30\,M_\odot$, the evolution after the MS is the following: a first crossing of the Hertzsprung-Russell diagram (HRD) up to the RSG branch, and then, due to the strong mass loss during the RSG phase, a second crossing occurs, the stars ending their life in the blue side of the HRD. There is thus a double population of blue supergiant stars (BSG): the first one consists in stars immediately after the MS that are in their first crossing \citep[group 1, see][]{Georgy2014a}, and the second one consists in stars that are post-RSG stars, that are going from the red side to the blue side of the HRD (group 2).

\section{Are $\alpha$ Cyg variables group 2 stars?}

The evolution in the HRD of post-MS stars in the range $20-30\,M_\odot$ computed in \citet{Ekstrom2012a} and \citet{Saio2013a} occurs at roughly constant luminosity. However, a major difference between the phase where the star is in group 1 and the phase in group 2 is the current mass. Indeed, the star encounters a very strong mass loss during the RSG phase, and is thus considerably less massive once it reaches again the BSG region. For example, a rotating model with an initial mass of $25\,M_\odot$ has a mass of $23.48\,M_\odot$ when it reaches for the first time $\log(T_\mathrm{eff}) = 4$ (group 1), and a mass of only $12.68\,M_\odot$ when it has the same $T_\mathrm{eff}$ during the second crossing (group 2). This makes the luminosity-to-mass ratio bigger for group 2 stars compared to group 1, and considerably changes the pulsationnal properties of these stars.

\begin{figure}
\begin{center}
\includegraphics[width=\textwidth]{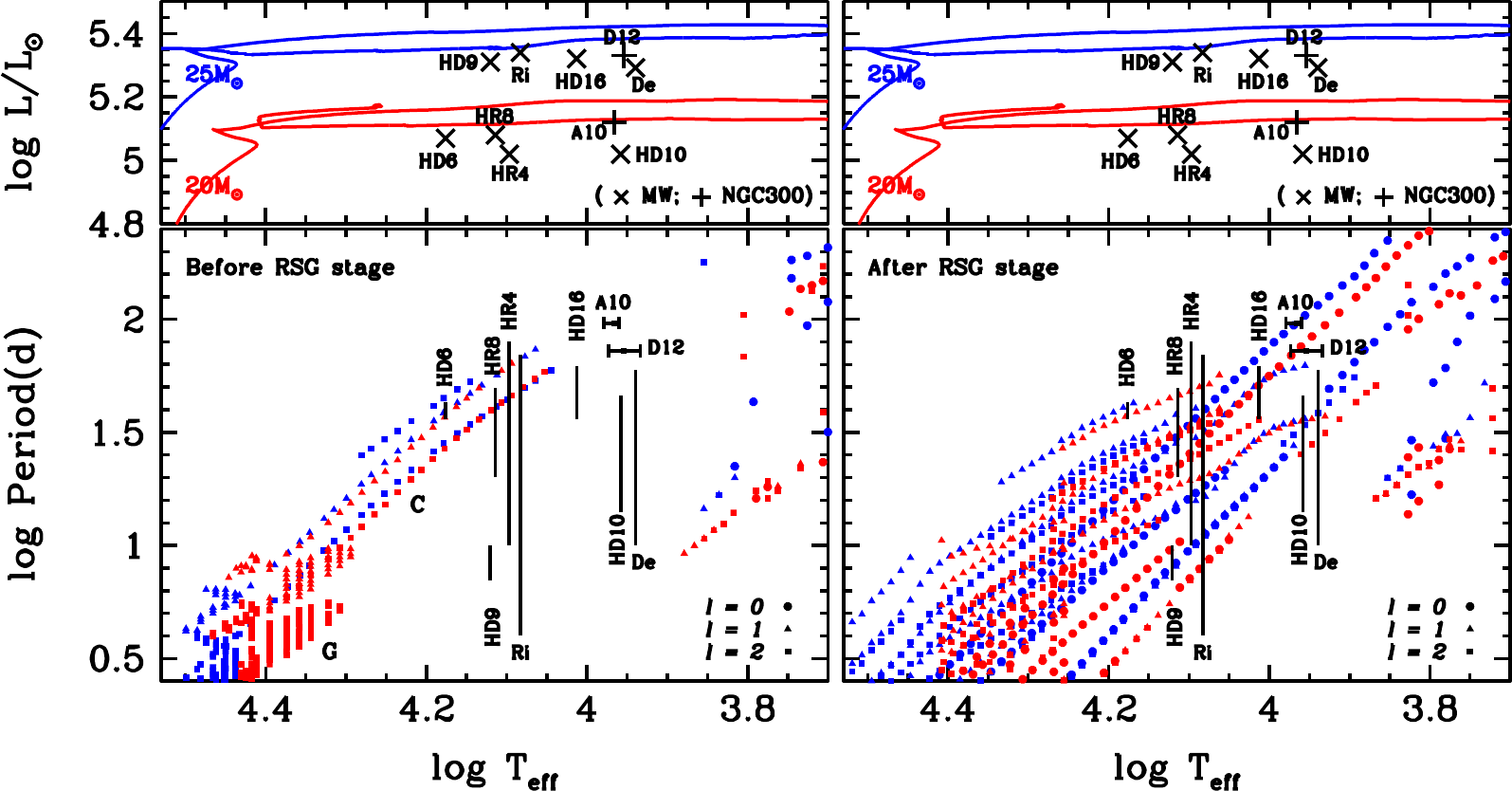}
\caption{Each side shows the tracks in the HRD \textit{(top panel)} and the pulsation periods of the excited modes \textit{(bottom panel)} of a rotating $20\,M_\odot$ (red) and $25\,M_\odot$ (blue) model at solar metallicity. \textit{Left panel:} Periods computed during the first crossing of the HRD (group 1). \textit{Right panel:} Periods computed during the second crossing of the HRD (group 2). Observational values are also indicated \citep{Firnstein2012a,Moravveji2012a,Leitherer1984a,vanLeeuwen1998a,Fraser2010b,Kaltcheva2010a,Kaufer1996a,Kaufer1997a,Sterken1977a,Sterken1999a,Schiller2008a,Richardson2011a,Markova2008a,Percy2008a,Kudritzki2008a,Bresolin2004a}. Figure adapted from \citet{Saio2013a}.}
\label{PulsaSch}
\end{center}
\end{figure}

Figure~\ref{PulsaSch} shows the period of the different excited modes for a group 1 model (left) and group 2 (model). These results indicate that in order to reproduce the observed period of $\alpha$ Cyg stars, it is unavoidable to lose a lot of mass to increase the $L/M$ ratio. This seems to indicate that these stars are post-MS stars and belong to group 2 rather than group 1.

\section{Surface abundances of group 2 stars}

For at least two $\alpha$ Cyg variables (Rigel and Deneb), some measurements of the surface abundances are available. In the following, we will focus on the N/C ratio. The observed ratio are the following \citep{Przybilla2010a}: $\mathrm{N}/\mathrm{C} = 2.0$ (Rigel) and $3.4$ (Deneb). These relatively small values indicate that the surface abundances of these stars are partially processed by the CNO cycle (the solar N/C ratio is $\sim 0.3$), but they are far from the values at the CNO equilibrium, $\sim 60$).

\begin{figure}
\begin{center}
\includegraphics[width=\textwidth]{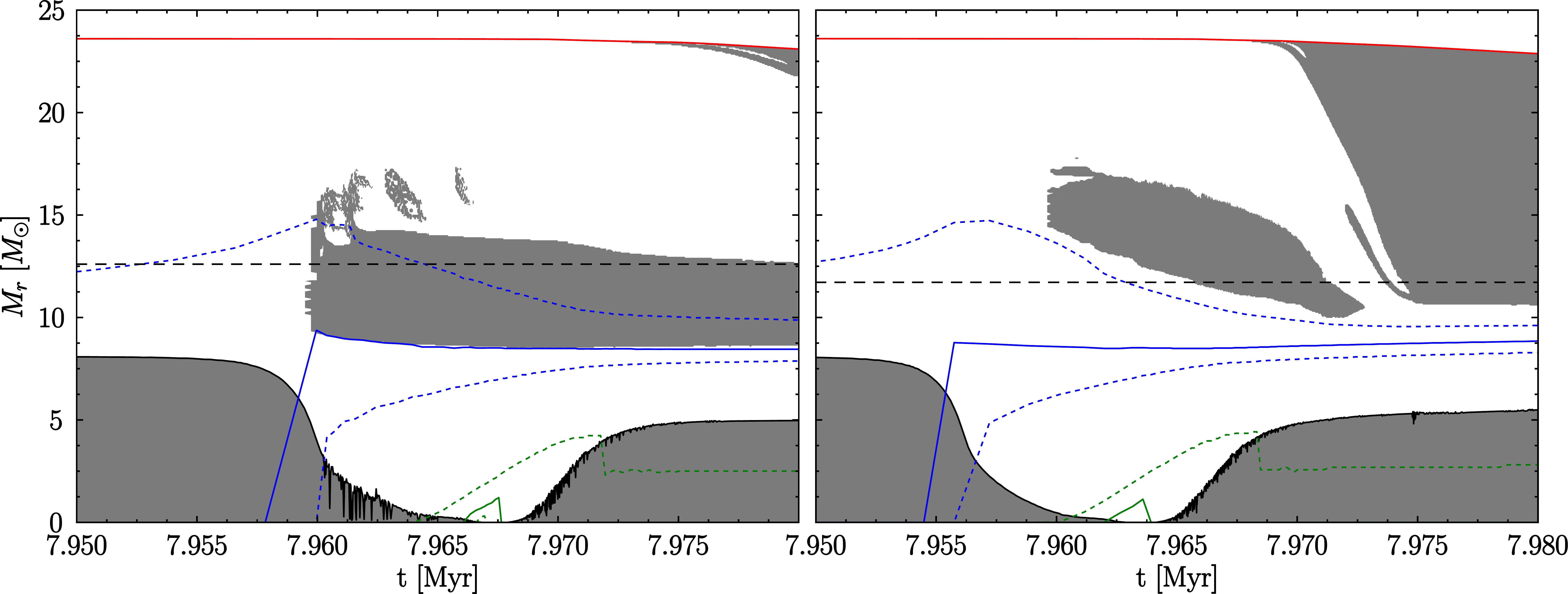}
\caption{Kippenhahn plot for a $20\,M_\odot$ model computed with the Schwarzschild criterion for convection \textit{(left panel)} and Ledoux criterion \textit{(right panel)}. Only the very end of the MS and the beginning of central He-burning is shown. Grey areas indicate the position of the convective zones. The red line shows the position of the stellar surface. Blue (green) solid line indicate the maximimum of the energy generation rate due to H-burning (He-burning) and the dotted line when the energy generation rate is $100\,\mathrm{erg}\,\mathrm{s}^{-1}\,\mathrm{g}^{-1}$. The black dashed line shows the layer that will be uncovered by mass loss when the star reaches $\log(T_\mathrm{eff}) = 4$ during the second crossing. Adapted from \citet{Georgy2014a}.}
\label{Kippen}
\end{center}
\end{figure}

The ``standard'' models shown above were computed using the Schwarzschild criterion for convection \citep[see][for the details of the physical processes implemented in these models]{Ekstrom2012a}. When the star becomes for the second time a BSG, and thus presents pulsation periods that are compatible with the observations of $\alpha$ Cyg variables, the surface N/C ratio is $58$, much more than the observed values. This is explained by looking at the Kippenhahn plot shown in Fig.~\ref{Kippen} \textit{(left panel)}. At the very end of the MS, when H-burning migrates from the centre to a shell, an important intermediate convective zone appears just on top of the previous convective core, bringing towards more external layers (up to a lagrangian coordinate $M_r \sim 15\,M_\odot$) material strongly processed by CNO cycle. Due to the strong mass loss the star will encounter during the RSG phase, these layers will be uncovered when the star enter in the group 2 region of the HRD, explaining the high N/C ratio.

As our model have to go through a previous RSG phase to explain the pulsationnal properties of $\alpha$ Cyg variables, the surface abundances of these stars tell us that something is missing in the treatment of internal mixing we used to compute these models.

\section{Model with the Ledoux criterion for convection}

As explained previously, the reason of the strong enrichment of the surface in our ``standard'' models is due to the position of the intermediate convective shell that appears at the ignition of the H-burning shell. A possible solution to this problem is thus to find a way to change the position of this convective zone. This can be achieved by changing the criterion that determines if a zone is convective or not. The Schwarzschild criterion, used in the ``standard models'', does not account for the stabilising effect of the presence of a chemical composition gradient. On the other hand, the Ledoux criterion account for this effect.

We have tried to compute a model with the Ledoux criterion instead of the Schwarzschild one \citep{Georgy2014a}. The corresponding Kippenhahn plot is shown in Fig.~\ref{Kippen} \textit{(right panel)}. As a consequence of the Ledoux criterion, the intermediate convective shell is now shifted towards the surface compared to the ``standard model'' \textit{(left panel)}, bringing towards the external layers of the star material that is less affected by the CNO cycle. The zone in the star that will be uncovered by the mass loss during the RSG phase has a C/N ratio of $\sim 7$, much closer to the observations than previously. In the same time, the pulsationnal properties of group 2 stars are still in good agreement with the observations.

Although the situation is improved, it is still not perfect. The surface gravity of Rigel and Deneb was also determined: $\log(g) = 1.75 \pm 0.1$ (Rigel), and $\log(g) = 1.20 \pm 0.1$ \citep[Deneb,][]{Przybilla2010a}. At effective temperatures corresponding to these stars, our models have $\log(g) = 1.56$ and $\log(g) = 0.76$ respectively. This quite big discrepancy is not solved to date, and further investigations are needed.

\section{Conclusions}

The main result of this work is to show how important is a good treatment of the convection in stellar evolution codes. It appeared that changing the criterion applied to determine if a zone is convective or not can drastically change the evolution of the surface properties of a stellar model. In that framework, the current development of observational technics such as asteroseismology combined with other methods (abundances measurements, ...) can bring new constraints for the stellar modeling.

On the theoretical/numerical side, some efforts have also to be made in order to improve our knowledge of convection, and improve how it is implemented in the classical 1d codes. The development of multi-d hydro-simulations of convection \citep[see e.g.][]{Meakin2007a,Viallet2013a} is an obvious step in that direction, and will probably lead to decisive changes in the way convection is treated in the numerical codes.

\bibliographystyle{iau307}
\bibliography{MyBiblio}

\begin{thebibliography}{}

\bibitem[\protect\astroncite{{Bresolin} et~al.}{2004}]{Bresolin2004a}
{Bresolin}, F., {Pietrzy{\'n}ski}, G., {Gieren}, W., {et~al.} 2004,
\newblock {\em \apj} 600, 182

\bibitem[\protect\astroncite{{de Jager} et~al.}{1988}]{deJager1988a}
{de Jager}, C., {Nieuwenhuijzen}, H., \& {van der Hucht}, K.~A. 1988,
\newblock {\em \aaps} 72, 259

\bibitem[\protect\astroncite{{Ekstr{\"o}m} et~al.}{2012}]{Ekstrom2012a}
{Ekstr{\"o}m}, S., {Georgy}, C., {Eggenberger}, P., {et~al.} 2012,
\newblock {\em \aap} 537, A146

\bibitem[\protect\astroncite{{Firnstein} \& {Przybilla}}{2012}]{Firnstein2012a}
{Firnstein}, M. \& {Przybilla}, N. 2012,
\newblock {\em \aap} 543, A80

\bibitem[\protect\astroncite{{Fraser} et~al.}{2010}]{Fraser2010b}
{Fraser}, M., {Dufton}, P.~L., {Hunter}, I., \& {Ryans}, R.~S.~I. 2010,
\newblock {\em \mnras} 404, 1306

\bibitem[\protect\astroncite{{Georgy}}{2012}]{Georgy2012a}
{Georgy}, C. 2012,
\newblock {\em \aap} 538, L8

\bibitem[\protect\astroncite{{Georgy} et~al.}{2013}]{Georgy2013b}
{Georgy}, C., {Ekstr{\"o}m}, S., {Eggenberger}, P., {et~al.} 2013,
\newblock {\em \aap} 558, A103

\bibitem[\protect\astroncite{{Georgy} et~al.}{2012}]{Georgy2012b}
{Georgy}, C., {Ekstr{\"o}m}, S., {Meynet}, G., {et~al.} 2012,
\newblock {\em \aap} 542, A29

\bibitem[\protect\astroncite{{Georgy} et~al.}{2014}]{Georgy2014a}
{Georgy}, C., {Saio}, H., \& {Meynet}, G. 2014,
\newblock {\em \mnras} 439, L6

\bibitem[\protect\astroncite{{Kaltcheva} \& {Scorcio}}{2010}]{Kaltcheva2010a}
{Kaltcheva}, N. \& {Scorcio}, M. 2010,
\newblock {\em \aap} 514, A59

\bibitem[\protect\astroncite{{Kaufer} et~al.}{1997}]{Kaufer1997a}
{Kaufer}, A., {Stahl}, O., {Wolf}, B., {et~al.} 1997,
\newblock {\em \aap} 320, 273

\bibitem[\protect\astroncite{{Kaufer} et~al.}{1996}]{Kaufer1996a}
{Kaufer}, A., {Stahl}, O., {Wolf}, B., {et~al.} 1996,
\newblock {\em \aap} 305, 887

\bibitem[\protect\astroncite{{Kudritzki} et~al.}{2008}]{Kudritzki2008a}
{Kudritzki}, R.-P., {Urbaneja}, M.~A., {Bresolin}, F., {et~al.} 2008,
\newblock {\em \apj} 681, 269

\bibitem[\protect\astroncite{{Leitherer} \& {Wolf}}{1984}]{Leitherer1984a}
{Leitherer}, C. \& {Wolf}, B. 1984,
\newblock {\em \aap} 132, 151

\bibitem[\protect\astroncite{{Markova} \& {Puls}}{2008}]{Markova2008a}
{Markova}, N. \& {Puls}, J. 2008,
\newblock {\em \aap} 478, 823

\bibitem[\protect\astroncite{{Martins} \& {Palacios}}{2013}]{Martins2013a}
{Martins}, F. \& {Palacios}, A. 2013,
\newblock {\em \aap} 560, A16

\bibitem[\protect\astroncite{{Meakin} \& {Arnett}}{2007}]{Meakin2007a}
{Meakin}, C.~A. \& {Arnett}, D. 2007,
\newblock {\em \apj} 667, 448

\bibitem[\protect\astroncite{{Moravveji} et~al.}{2012}]{Moravveji2012a}
{Moravveji}, E., {Guinan}, E.~F., {Shultz}, M., {Williamson}, M.~H., \& {Moya},
  A. 2012,
\newblock {\em \apj} 747, 108

\bibitem[\protect\astroncite{{Percy} et~al.}{2008}]{Percy2008a}
{Percy}, J.~R., {Palaniappan}, R., {Seneviratne}, R., {Adelman}, S.~J., \&
  {Markova}, N. 2008,
\newblock {\em \pasp} 120, 311

\bibitem[\protect\astroncite{{Przybilla} et~al.}{2010}]{Przybilla2010a}
{Przybilla}, N., {Firnstein}, M., {Nieva}, M.~F., {Meynet}, G., \& {Maeder}, A.
  2010,
\newblock {\em \aap} 517, A38+

\bibitem[\protect\astroncite{{Richardson} et~al.}{2011}]{Richardson2011a}
{Richardson}, N.~D., {Morrison}, N.~D., {Kryukova}, E.~E., \& {Adelman}, S.~J.
  2011,
\newblock {\em \aj} 141, 17

\bibitem[\protect\astroncite{{Saio} et~al.}{2013}]{Saio2013a}
{Saio}, H., {Georgy}, C., \& {Meynet}, G. 2013,
\newblock {\em \mnras} 433, 1246

\bibitem[\protect\astroncite{{Schiller} \& {Przybilla}}{2008}]{Schiller2008a}
{Schiller}, F. \& {Przybilla}, N. 2008,
\newblock {\em \aap} 479, 849

\bibitem[\protect\astroncite{{Sterken}}{1977}]{Sterken1977a}
{Sterken}, C. 1977,
\newblock {\em \aap} 57, 361

\bibitem[\protect\astroncite{{Sterken} et~al.}{1999}]{Sterken1999a}
{Sterken}, C., {Arentoft}, T., {Duerbeck}, H.~W., \& {Brogt}, E. 1999,
\newblock {\em \aap} 349, 532

\bibitem[\protect\astroncite{{van Leeuwen} et~al.}{1998}]{vanLeeuwen1998a}
{van Leeuwen}, F., {van Genderen}, A.~M., \& {Zegelaar}, I. 1998,
\newblock {\em \aaps} 128, 117

\bibitem[\protect\astroncite{{Vanbeveren} et~al.}{1998}]{Vanbeveren1998a}
{Vanbeveren}, D., {De Donder}, E., {van Bever}, J., {van Rensbergen}, W., \&
  {De Loore}, C. 1998,
\newblock {\em \na} 3, 443

\bibitem[\protect\astroncite{{Viallet} et~al.}{2013}]{Viallet2013a}
{Viallet}, M., {Meakin}, C., {Arnett}, D., \& {Moc{\'a}k}, M. 2013,
\newblock {\em \apj} 769, 1

\end{thebibliography}

\begin{discussion}

\discuss{Noels}{Comment: In models computed with the Schwarzschild criterion, the layers in the $\mu$-gradient region become convective while they should be semi-convective. This would lower the efficiency of the mixing and would lower the ratios N/C and N/O.}

\discuss{Noels}{What would happen to the ratio if no intermediate convective zone at all is present? It would still be possible to affect the N/C and N/O ratios if mass loss in the RSG phase is large enough to uncover the layers where CNO cycle has affected the CNO abundances already in the early phases of the main sequence.}

\discuss{Georgy}{It has to be checked. As the mass loss uncover quite deep layers, it probably shows some material affected at least partially by CNO cycle. This could be easily checked from the structure at the end of the MS, before the intermediate convective zone develops.}

\discuss{Moravveji}{Where does all lost hydrogen rich envelope go during the RSG phase? Do we have observations for that?}

\discuss{Georgy}{Some indirect observations of interactions with a dense circumstellar shell exist. For example, type IIn supernovae, or the presence of ring nebulae around Wolf-Rayet stars.}

\discuss{Przibilla}{How do the lifetimes of blue supergiants on the first crossing of the HRD to the red and those of the second crossing towards the blue compare?}

\discuss{Georgy}{It is extremely dependent on the stellar model (e.g. rotation or not) and can change from 10\%-90\%. So far, it seems to be difficult to assess firm statistics on that point.}

\end{discussion}

\end{document}